\documentclass[10pt,twocolumn,a4paper]{IEEEtran}

\usepackage[utf8]{inputenc}
\usepackage{graphicx}
\usepackage[T1]{fontenc}
\usepackage{url}
\usepackage{ifthen}
\usepackage[cmex10]{amsmath} % Use the [cmex10] option 
\usepackage{amsmath}
\usepackage{setspace,color}
\usepackage{amssymb}
\usepackage{multicol}

\makeatletter
\newtheorem{definition}{Definition}
\newtheorem{theorem}{Theorem}
\newtheorem{proposition}{Proposition}
\newtheorem{lemma}{Lemma}
\newtheorem{example}{Example}
\newtheorem{corollary}{Corollary}
\newtheorem{remark}{Remark}
\makeatother

\def\a{{\alpha}}

\newcommand{\defined}{\triangleq}

\newcommand{\graph}{\set{G}}
\newcommand{\nodes}{\set{V}}
\newcommand{\edges}{\set{E}}
\newcommand{\N}{\set{N}}

\newcommand{\X}{\set{X}}
\newcommand{\A}{\set{A}}

\begin{document}

\newcommand{\seq}[3]{{#1_{#2}, \ldots, #1_{#3}}}
\newcommand{\newsym}[1]{#1}
\newcommand{\newnot}[1]{#1}
\def\defined{\triangleq}
\def\sessions{{\cal S}}

\title{Characterising Probability Distributions via Entropies}
\author{\IEEEauthorblockN{Satyajit Thakor$^{\dagger}$,~Terence Chan$^{\ddagger}$ and Alex Grant$^{*}$}\\
\IEEEauthorblockN{Indian Institute of Technology Mandi$^{\dagger}$}\\
%School of Computing and Electrical Engineering$^{\dagger}$}\\
\IEEEauthorblockN{University of South Australia$^{\ddagger}$} \\
\IEEEauthorblockN{Myriota Pty Ltd$^{*}$} \\
}
\maketitle

\begin{abstract}
Characterising the capacity region for a network can be extremely difficult, especially when the sources are dependent. Most existing computable outer bounds are relaxations of the Linear Programming bound. 
One main challenge to  extend linear program bounds to the case of  correlated sources  is the difficulty (or impossibility) of characterising arbitrary dependencies via entropy functions. 
This paper tackles the problem by addressing how to use  entropy functions to characterise correlation among sources. 
% We show that by using carefully chosen auxiliary random variables, the characterisation can be fairly ``accurate''. 
\end{abstract}

\newcommand{\dist}[1]{{\Omega}_{#1}}
\newcommand{\setindex}[1]{{ {\mathcal P}_{[2,#1]} }}

\def\lr{{\langle}}
\def\rr{{\rangle}}
\def\graph{{\mathcal G}}
\def\nodes{{\mathcal V}}
\def\edges{{\mathcal E}}
\newcommand\eht[2]{{#1 \to #2}}
\def\A{{\mathcal A}}
%%%%%%%%%%%%%%%%%%%%%%%%%%%%%%%%%%%%%%%%%%%%%

\section{Introduction}

This paper begins with a very simple and well known result.  
Consider a binary random variable $X$ such that 
\begin{align*}
%p_{X}(0) & = p \\
%p_{X}(1) & = 1-p.
p_{X}(0)  = p \text{ and } p_{X}(1)  = 1-p.
\end{align*}
While the entropy of $X$ does not determine exactly what the probabilities of $X$ are, it essentially determines the probability distribution (up to relabelling).  To be precise, let $0\le q \le 1/2$  such that 
$
H(X) = h_{b}(q)
$
where
$$
h_{b}(q) \triangleq -q\log q - (1-q) \log (1-q).
$$
Then 
either $p=q$ or $p= 1-q$. Furthermore, the two possible distributions can be obtained from each other by renaming the random variable outcomes appropriately. 
In other words,  there is a  one-to-one correspondence between entropies and distribution (when the random variable is binary).

The basic question now is: \emph{How ``accurate'' can entropies specify the distribution of random variables}? When $X$ is not binary, the entropy $H(X)$ alone is not sufficient to characterise the probability distribution of $X$. In \cite{ThaChaGra13}, it was proved that if $X$ is a random scalar variable, its distribution can still be determined by using auxiliary random variables \emph{subject to alphabet cardinality constraint}. The results can also be extended to random vector if the distribution is positive. 
However, the proposed approach cannot be generalised to the case when the distribution is not positive. 
In this paper, we take a different approach and generalise the result to any random vectors.
Before we continue answering the question, we will briefly describe an application (based on network coding problems) of characterising distributions (and correlations) among random variables by using entropies.   
 
Let the directed acyclic graph $\graph = (\nodes, \edges)$ serve as a simplified model of a communication \emph{network} with error-free point-to-point communication links. Edges $e\in\edges$ have finite capacity $C_e>0$. 
Let $\sessions$  be an index set for a number of multicast sessions, and  $\{Y_{s}: s \in \sessions\}$  be the set of source random variables. These sources are available at the nodes identified by the mapping (a source may be available at multiple nodes)
$
a : {\sessions} \mapsto 2^{\nodes}
$.
Similarly, each source may be demanded by multiple sink nodes, identified by the mapping
$
b : \sessions \mapsto 2^{\mathcal V}
$. 
For all $s$ assume that $a(s)\cap b(s)=\emptyset$. Each edge $e\in\edges$ carries a random variable $U_e$  which is a function of incident edge random variables and source random variables.

Sources are i.i.d. sequences
$
\{(Y_{s}^{n}, s\in\sessions) , \: n = 1, 2, \ldots, \}. 
$
Hence, each $(Y_{s}^{n}, s\in\sessions)$ has the same joint distribution, and is independent across different $n$. For notation simplicity, we will use  $(Y_{s}, s\in\sessions)$ to denote a generic copy of the sources at any particular time instance.
However, within the same ``time'' instance $n$, the random variables $(Y_{s}^{n}, s\in\sessions)$ may be correlated.
We assume that the distribution of $(Y_{s}, s\in\sessions)$ is known. 
 
Roughly speaking, a link capacity tuple $\mathbf{C} = (C_{e}: e \in \mathcal E)$ is achievable if one can design a network coding solution to transmit the sources $
\{(Y_{s}^{n}, s\in\sessions) , \: n = 1, 2, \ldots, \}$ to their respective destinations such that 1) the probability of decoding error is vanishing (as $n$ goes to infinity), and 2) the number of bits transmitted on the link $e\in \mathcal E$ is at most $n C_{e}$.
The set of all achievable link capacity tuples is denoted by $\mathcal R$.

\begin{theorem}[Outer bound~\cite{ThaChaGra11}]\label{thm2}
For a given network, consider the set of correlated sources 
$
(Y_s, s\in {\mathcal S})
$ 
with underlying probability distribution $P_{Y_{\mathcal S}}(\cdot)$. Construct any auxiliary random variables
$
(K_{i},i\in \mathcal L)
$ 
by choosing a conditional probability distribution function
$P_{K_{\mathcal L}|Y_{\mathcal S}}(\cdot)$. 
Let $\mathcal R' $ be the set of all link capacity tuples 
$\mathbf{{C}} = (C_{e}: e \in \mathcal E)$
 such that there exists a polymatroid $ h $ satisfying the following constraints
\begin{align}
h (X_{\mathcal W}, J_{\mathcal Z})-H(Y_{\mathcal W},K_{\mathcal Z})&=0\\
%&\forall \mathcal W \subseteq \mathcal S, \mathcal Z \subseteq \mathcal L\nonumber\\
h (U_{e}|X_{s}: a(s) \rightarrow e,U_{f}: f \rightarrow e)&=0\label{eq:Encoding}\\
h (Y_{s}:u \in b(s)| X_{s'}: u \in a(s'), U_{e}: e \rightarrow u)&=0\label{eq:Decoding}\\
%&\forall u \in b(s) \nonumber, s \in \mathcal S\\
 C_{e} -h (U_{e})&\geq 0 \label{eq:EdgeCapacity}
\end{align}
for all $\mathcal W \subseteq \mathcal S, \mathcal Z \subseteq \mathcal L, e\in \mathcal E, u \in b(s)$ and $s\in\mathcal S$.
Then
\begin{equation}
\mathcal R \subseteq \mathcal R'
\end{equation}
where the notation $x\rightarrow y$ means $x$ is incident to $y$ and $x,y$ can be an edge or a node.
\end{theorem}
%\begin{IEEEproof}[Sketch of Proof]
%The improved outer bound is an ordinary LP bound obtained by  viewing the auxiliary random variables as ``virtual sources'' which are not available at and not demanded by any nodes. 
%%In that case, the resulting LP bound  
%%
%% follows since if we choose degenerate auxiliary random variables or if $\mathcal L$ is empty set then $\mathcal R'(\overline{\Gamma^*})$ and $\mathcal R'(\Gamma)$ reduce to $\mathcal R(\overline{\Gamma^*})$ and $\mathcal R(\Gamma)$ respectively. On the other hand, any other choice of auxiliary random variables cannot increase the regions $\mathcal R(\overline{\Gamma^*})$ and $\mathcal R(\Gamma)$ (i.e., can only decrease or keep the regions same).
%\end{IEEEproof}
%
% 

\begin{remark}
The region $R' $ will depend on how we choose the auxiliary random variables 
$
(K_{i},i\in \mathcal L)
$. In the following, we give an example to illustrate this fact. 
\end{remark}

\begin{figure}[htbp]
\centering
  \includegraphics[scale=0.35]{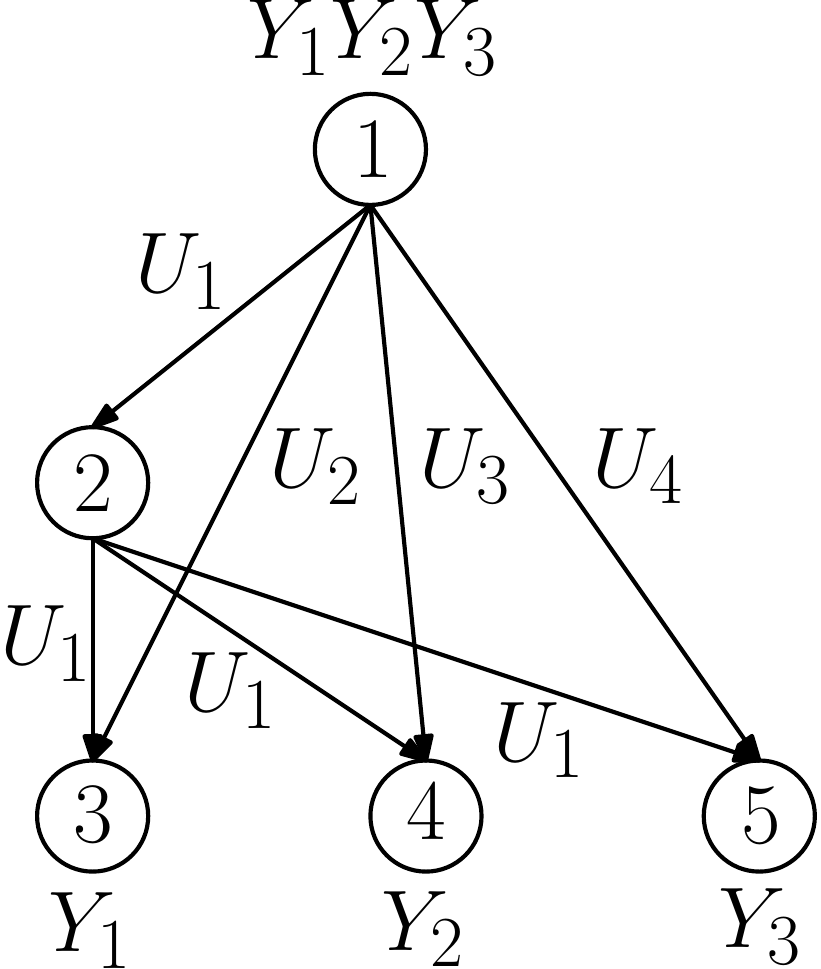}
  \caption{A network example \cite{ThaChaGra11}.}\label{fig:net-cs}
\end{figure}

Consider the following network coding problem depicted in Figure \ref{fig:net-cs}, in which three correlated sources $Y_1, Y_2, Y_3$ are available at node 1 and are demanded at nodes $3,4,5$ respectively. Here, $Y_1, Y_2, Y_3$ are defined such that 
$Y_1 = (b_0,b_1)$, $Y_2 = (b_0,b_2)$ and $Y_3 = (b_1,b_2)$ for some independent and uniformly distributed binary random variables  $b_0,b_1,b_2$. Furthermore, the edges from node $2$ to nodes $3,4,5$ have sufficient capacity to carry the random
variable $U_1$ available at node 2. 

%\begin{proposition}
We consider two outer bounds obtained from Theorem \ref{thm2} for the above network coding problem. In the first scenario, we use no auxiliary random variables, while in the second scenario, 
we use three auxiliary random variables such that 
\[
K_{0} = b_{0}, \: K_{1} = b_{1}, \: K_{2} = b_{2}.
\] 
Let $\mathcal R_{i}$ be respectively the outer bounds for the  two scenarios. Then $\mathcal R_{2}$ is a proper subset of $\mathcal R_{1}$.  In particular, the link capacity tuple 
$(C_e=1, e=1,...,4)$
 is in the region $\mathcal R_{1} \setminus \mathcal R_{2}$ \cite{ThaChaGra11}. 
%\end{proposition}
This example shows that by properly choosing auxiliary random variables, one can better capture the correlations among the sources, leading to a strictly tighter/better outer bound for network coding. Construction of auxiliary random variables from source correlation was also considered in \cite{GohYanJagg13} to improve cut-set bounds.

\section{Main results}
%%%%%%%%%%%%%%%%%%%%%%%%%%%%%%%%%%%%%
%In previous section, we have seen that correlation among sources can be better characterised by using auxiliary random variables. 
%
%%
%To illustrate the idea, consider a random vector   $X=(X_{1}, \ldots, X_{M})$ with probability distribution $p_{X} (x_{1},\ldots, x_{M})$.  We can arbitrarily ``construct'' an auxiliary random variable $Y$ by specifying the conditional probability distribution $p_{Y|X} (y|x_{1},\ldots, x_{M})$. Now, instead of using only the entropy function defined in \eqref{1}, we can improve the ``quality'' of representation by using the ``extended entropy function'' 
%\begin{align}
%h(W) \triangleq 
%\begin{cases}
%H(X_{s}  , s\in\alpha) & \text{ if } Y \not\in \alpha \\
%H(Y, X_{s}  , s\in\alpha) & \text{ if } Y \in \alpha \\
%\end{cases}
%\end{align}
%for all subset of random variables  $W \subseteq \{X_{1}, \ldots X_{M}, Y\}$. 
%
%For example, suppose one can construct an auxiliary random variable $Y$ such that 
%\begin{align}\label{3}
%H(Y|X_{1}) = H(Y|X_{2}) = 0 
%\end{align} 
%and 
%\begin{align}\label{4}
%H(Y) \ge \frac{1}{2} \max ( H(X_{1}), H(X_{2)}).
%\end{align}
%The conditions \eqref{3} and \eqref{4} already impose a very strong constraint on the joint probability distribution of $(X_{1},  X_{2})$ that 
%$X_{1}$ and $X_{2}$ have a ``common information'' $Y$ of entropy at least half of the entropy of  each individual random variable.
%

\def\PN{\{2,\ldots, m \}}

\def\X{{\cal X}}
\def\a{{\alpha}}

%%%%%%%%%%%%%%%%%%%%%%%%%%%%%%%%%%%
%\subsection{Properties of partition induced random variables}

In this section, we will show that by using auxiliary random variables, the probability distribution of a set of random variables (or a random vector) can be uniquely characterised from the entropies of these variables.

\subsection{Random Scalar Case}\label{sec:AuxVariables}

Consider any ternary random variable $X$. Clearly, entropies of $X$ and probability distributions are not in one-to-one correspondence. 
In \cite{ThaChaGra13}, auxiliary random variables are used to in order to exactly characterise the distribution.  

%The idea is best demonstrated by an example. 
Suppose $X$ is ternary, taking values from the set $\{1,2,3\}$. Suppose also that $p_{X}(x) > 0$ for all $x\in \{1,2,3\}$.
Define random variables $A_{1}$, $A_{2}$ and $A_{3}$ such that 
\begin{align}\label{2b}
A_{i} = 
\begin{cases}
1 & \text{ if } X = i \\
0 & \text{ otherwise. }
\end{cases}
\end{align}
Clearly, 
\begin{align}
H(A_{i}|X) &= 0,\label{2a}\\% \text{ and } \label{2a}\\
%\end{align}
%\begin{align}
H(A_{i}) &= h_{b}(p_{X}(i)).\label{2}
\end{align}
Let us further assume that $p_{X}(i) \le 1/2$ for all $i$. Then by \eqref{2} and strict monotonicity of 
$h_{b}(q)$ in the interval $[0, 1/2]$, it seems at the first glance that the distribution of $X$ is uniquely specified by the entropies of the auxiliary random variables. 

However,  there is a catch in the argument -- The auxiliary random variables chosen are not arbitrary. When we ``compute'' the probabilities of $X$ from the entropies of the auxiliary random variables, it is assumed to know how the random variables are constructed. Without knowing the ``construction'',  it is unclear how to find the probabilities of $X$ from entropies.

More precisely, suppose we only know that there exists auxiliary random variables $A_{1},A_{2},A_{3}$ such that \eqref{2a} and \eqref{2} hold (without knowing that the random variables are specified by \eqref{2b}). Then we cannot determine precisely what the distribution of $X$ is.  
Despite this complexity, \cite{ThaChaGra13,ThaChaGra11}  showed  a construction of auxiliary random variables from which the probability distribution can be characterised from entropies. The results will also be briefly restated as a necessary prerequisite for the vector case.

Let $X$ be a random variable  with support ${\cal N}_{n}=\{1,\ldots,n\}$ and $\Omega$ be the set of all nonempty binary partitions of ${\cal N}_{n}$. In other words, $\Omega$ is 
the collection of all sets $\{\alpha, \alpha^{c}\}$ such that $\alpha \subseteq {\cal N}_{n}$,  and both $|\alpha| $ and $|\alpha^{c}|$ are nonzero.
We will use $\lr \alpha \rr$ to denote the set $\{\alpha, \alpha^{c}\}$. 
To simplify notations, we may assume without loss of generality that $  \alpha $ is  a subset of $  \{2,\ldots,n\} $.
 Clearly, $|\Omega| = 2^{n-1}-1$.
Unless explicitly stated otherwise,  we may assume without loss of generality that 
 the probability that $X=i$ (denoted by $p_i$) is monotonic decreasing. In other words, 
$$p_1 \geq \ldots \geq p_n>0.$$

\begin{definition}[Partition Random Variables]\label{def:RVsAalpha}
A random variable $X$ with support  ${\cal N}_{n}$  induces $2^{n-1}-1$ random variables
$A_{\lr \alpha \rr}$ for  $\alpha \in \Omega$ such that 
\begin{equation}
  A_{\lr \alpha \rr} \triangleq \left\{
  \begin{array}{l l}
    \alpha & \quad \text{if $X \in \alpha$}\\
    \alpha^{c} & \quad \text{otherwise.}\\
  \end{array} \right.
\end{equation}
%
%
%
%\begin{equation}
%  A_{\lr \alpha \rr} \triangleq \left\{
%  \begin{array}{l l}
%    1 & \quad \text{if $X \in \alpha$}\\
%    0 & \quad \text{otherwise.} 
%  \end{array} \right.
%\end{equation}
We called $\{A_{\lr \alpha \rr}, \alpha \in \Omega \}$ the collection of \emph{binary partition random variables} of $X$.
\end{definition}

%\begin{remark}
%Sometimes, it is instrumental to rename the partition random variables such that  
%\begin{equation}
%  A_{\lr \alpha \rr} \triangleq \left\{
%  \begin{array}{l l}
%    \alpha & \quad \text{if $X \in \alpha$}\\
%    \alpha^{c} & \quad \text{otherwise.}\\
%  \end{array} \right.
%\end{equation}
%\end{remark}

\begin{remark}
If $|\alpha| = 1 $ or $n-1$,   then there exists an element $i\in \X$ such that $A_{\lr \alpha \rr} = \{i\}$ if and only if $X= i$. Hence, $A_{\lr \alpha \rr}$ is essentially a binary variable \emph{indicating/detecting whether $X = i$ or not}. As such, we call $A_{\lr \alpha \rr}$ an \emph{indicator variable}. 
Furthermore, when $n\ge 3$, there are exactly $n$ indicator variables, one for each element in ${\cal N}_{n}$.
\end{remark}

%Naturally, if we are given the entropy of the indicator random variable $H(A_{\lr i \rr})$, then we can immediately determine what the probability $p_i$ is. In particular, if we further know that 
%\[
%p_i \le 1/2
%\]
%for all $i\ge 2$. Then $p_i$ is the unique value between [0,1/2] such that 
%\[
%h_{b}(p_{i}) = H(A_{\lr i \rr}).
%\]
%Also, $p_{1} = 1 - \sum_{i\ge  2} p_{i}$.
%

\begin{theorem}[Random Scalar Case]\label{thm1}
Suppose $X$ is a random variable with support $\N_{n}$.   
For any $\lr\alpha\rr \in \Omega$, let $A_{\lr \alpha \rr}$ be the corresponding binary partition random variables. 
Now, suppose  $X^{*}$ is another  random variable  such that 1) the size of its support $\X^{*}$ is at most the same as that of $X$, and 2) there exists random variables $(B_{\lr \alpha \rr}, \alpha \in\Omega)$ satisfying the following conditions:
\begin{align}
H(B_{\lr \alpha \rr}, \alpha \in \Delta) & = H(A_{\lr \alpha \rr},\alpha\in\Delta) \label{eq34}\\
H( B_{\lr \alpha \rr} | X^{*}) & =0 \label{eq35}
\end{align}
for all $\Delta \subseteq \Omega$.  Then
there is a mapping 
$$
\sigma : \N_{n} \to \X^{*}
$$
such that 
$
\Pr(X=i) = \Pr(X^{*} = \sigma(i)).
$
In other words, the probability distributions of $X$ and $X^{*}$ are essentially the same (via renaming outcomes).
\end{theorem}
\begin{IEEEproof}
A sketch of the proof is shown in Appendix \ref{app.a}.
\end{IEEEproof}

%
%The probability distributions of two scalar random variables are deemed equivalent if the sequences of positive probability masses (when arranged in descending order) are the same for both distributions. However, 
%
%one cannot naturally generalise the condition to vector variables. 
\subsection{Random Vector Case}

Extension of Theorem \ref{thm1} to the case of random vector has also been considered briefly in our previous work \cite{ThaChaGra13}. However, the extension is fairly limited in that work -- the random vector must have a positive probability distribution and each individual random variable must take at least three possible values. In this paper, we overcome these restrictions and fully generalise Theorem \ref{thm1} to the random vector case. 
%First, we begin with an example to illustrate a key difference between the scalar and the vector case. 

\begin{example}
Consider two random vectors  $X =(X_{1}, X_{2})$ and $X^{*} =(X^{*}_{1}, X^{*}_{2})$ with probability distributions given in Table \ref{table1}.
%%%%%%%%%%%%%%
\begin{table}[]
\centering
\caption{Probability distributions of $X$ and $X^{*}$}
\label{table1}
\begin{tabular}{cccccc}
  &                        &                          & $X_{2}$                        &                          &                          \\
  &                        & $1$                        & $2$                        & $3$                        & $4$                        \\ \cline{3-6} 
  & \multicolumn{1}{c|}{$a$} & \multicolumn{1}{c|}{$1/8$} & \multicolumn{1}{c|}{$1/8$} & \multicolumn{1}{c|}{$0$}    & \multicolumn{1}{c|}{$0$}    \\ \cline{3-6} 
$X_{1}$ & \multicolumn{1}{c|}{$b$} & \multicolumn{1}{c|}{$1/8$} & \multicolumn{1}{c|}{$1/8$} & \multicolumn{1}{c|}{$0$}    & \multicolumn{1}{c|}{$0$}    \\ \cline{3-6} 
  & \multicolumn{1}{c|}{$c$} & \multicolumn{1}{c|}{$0$}    & \multicolumn{1}{c|}{$0$}    & \multicolumn{1}{c|}{$1/8$} & \multicolumn{1}{c|}{$1/8$} \\ \cline{3-6} 
  & \multicolumn{1}{c|}{$d$} & \multicolumn{1}{c|}{$0$}    & \multicolumn{1}{c|}{$0$}    & \multicolumn{1}{c|}{$1/8$} & \multicolumn{1}{c|}{$1/8$} \\ \cline{3-6} 
%\end{tabular}
%\end{table}
%
&&&&\\
%&&&&\\
%&&&&\\
%
%\begin{table}[]
%\centering
%\caption{Probability distribution for $X=(X_{1},X_{2})$}
%\label{my-label}
%\begin{tabular}{llllll}
  &                        &                          & $X^{*}_{2}$                        &                          &                          \\
  &                        & $1$                        & $2$                        & $3$                        & $4$                        \\ \cline{3-6} 
  & \multicolumn{1}{c|}{$a$} & \multicolumn{1}{c|}{$1/8$} & \multicolumn{1}{c|}{$1/8$} & \multicolumn{1}{c|}{$0$}    & \multicolumn{1}{c|}{$0$}    \\ \cline{3-6}
$X^{*}_{1}$ & \multicolumn{1}{c|}{$b$} & \multicolumn{1}{c|}{$0$} & \multicolumn{1}{c|}{$1/8$} & \multicolumn{1}{c|}{$1/8$}    & \multicolumn{1}{c|}{0}    \\ \cline{3-6} 
  & \multicolumn{1}{c|}{$c$} & \multicolumn{1}{c|}{$0$}    & \multicolumn{1}{c|}{$0$}    & \multicolumn{1}{c|}{$1/8$} & \multicolumn{1}{c|}{$1/8$} \\ \cline{3-6} 
  & \multicolumn{1}{c|}{$d$} & \multicolumn{1}{c|}{$1/8$}    & \multicolumn{1}{c|}{$0$}    & \multicolumn{1}{c|}{$0$} & \multicolumn{1}{c|}{$1/8$} \\ \cline{3-6} 
\end{tabular}
\end{table}
%%%%%%%%
If we compare the joint probability distributions of 
$X $ and $X^{*}$, they are different from each other. Yet, if we treat  $X $ and $ X^{*} $ as scalars (by properly renaming), then they indeed have the same distribution (both uniformly distributed over a support of size 8). 
This example shows that we cannot directly apply Theorem \ref{thm1} to the random vector case, by simply mapping a vector into a scalar.

%
%
%their sequences of positive probability mass of $(X_{1},X_{2})$ and $(X^{*}_{1},X^{*}_{2})$ are indeed the same. The same is true for $X_{i}$ and $X^{*}_{i}$ for $i=1,2$. 
\end{example}

%%%%%%%%%%%%%%%%%%%%%%%%%%%%%%%%%%%%%%%
\begin{theorem}[Random Vector]\label{thm4}
Suppose $X=(X_{1},\ldots, X_{n})$ is a random vector with support $\X $  of size at least 3. 
Again, let $\Omega$ be the set of all nonempty binary partitions of $\X$ and  $A_{\lr \alpha \rr}$ be the binary partition random variable of $X$ such that 
\begin{equation}
  A_{\lr \alpha \rr} = \left\{
  \begin{array}{l l}
    \alpha & \quad \text{if $X \in \a$}\\
    \alpha^{c} & \quad \text{otherwise} 
  \end{array} \right.
\end{equation}
 for all $\lr \alpha \rr \in \Omega$.

Now, suppose  $X^{*} = (X^{*}_{1} , \ldots, X^{*}_{M})$ is another random vector  where  there exists random variables 
$$
(B_{\lr \alpha \rr}, \lr\alpha\rr \in\Omega)
$$ 
such that for any subset $\Delta$ of $\Omega$ and $\tau \subseteq \{1, \ldots, M\}$,
\begin{multline}
H(B_{\lr \alpha \rr}, \lr\alpha\rr \in \Delta, X^{*}_{j}, j \in \tau)  \\
= H(A_{\lr \alpha \rr}, \lr\alpha\rr \in \Delta, X_{j}, j \in \tau). 
\end{multline}
 Then the joint probability distributions of $X=(X_{1},\ldots, X_{n})$ and $X^{*} = (X^{*}_{1}, \ldots, X^{*}_{n})$ are essentially the same. More precisely,  there exists bijective mappings $\sigma_{m}$  for $m=1,\ldots, M$ such that 
\begin{multline}
\Pr (X = (x_{1}, \ldots , x_{M}) ) \\
= \Pr (X^{*} = (\sigma_{1}(x_{1}), \ldots , \sigma_{M}(x_{M}) )). 
\end{multline}
\end{theorem}
\begin{IEEEproof}
See Appendix B.%\ref{app.C}.
\end{IEEEproof}

\subsection{Applications: Network coding outer bounds} 
Together with Theorem \ref{thm2} and the  characterisation of  random variable using entropies, we obtain the following  outer bound $\mathcal R'' $   on the set of achievable capacity tuples.

\begin{corollary}\label{def:Improved Outer Bound-vector}
For any given network, consider the set of correlated sources $(Y_s, s\in {\mathcal S})$ with underlying probability distribution $P_{Y_{\mathcal S}}(\cdot)$. From this distribution, construct binary partition random variables
$A^{\mathcal W}_{\langle \alpha \rangle}$ for every subset $\mathcal W \subseteq \sessions$ as described in Theorem \ref{thm2} (for scalar subsets) and Theorem \ref{thm4} (for vector subsets). 
Let $\mathcal R''$ be the set of all link capacity tuples $\mathbf{{C}} = (C_{e}: e \in \mathcal E)$ such that there exists an almost entropic function $ h  \in \overline{\Gamma^*}$ satisfying the constraints \eqref{eq:Encoding}-\eqref{eq:EdgeCapacity} and 
%For any
%for given source random
%variables $Y_{\mathcal S} \sim P_{Y_{\mathcal S}}(\cdot)$ and auxiliary random variables $K_{\mathcal L} \sim P_{K_{\mathcal L}|Y_{\mathcal S}}(\cdot)$ which supports the following constraints for all
\begin{align}
h (X_{\mathcal W}, B^{\mathcal W}_{\langle \alpha \rangle})-H(Y_{\mathcal W},A^{\mathcal W}_{\langle \alpha \rangle})&=0%\\
%&\forall \ \mathcal W \subseteq \mathcal S, \langle \alpha \rangle \in \Omega  \\
% h (U_{e}|\{Y_{s}: a(s) \rightarrow e\},\{U_{f}: f \rightarrow e\}) &= 0\\
 %h (Y_{s}:u \in b(s)| Y_{s'}: s' \rightarrow u, U_{e}: e \rightarrow u) &=0\\
% &\forall u \in b(s), s \in \mathcal S\\
 %C_{e} -h (U_{e}) &\geq 0
%H(Y_{\mathcal W},A_{\langle \alpha \rangle})&= h (X_{\mathcal W}, B_{\langle \alpha \rangle})\\%,\forall \ \mathcal W \subseteq \mathcal S, \langle \alpha \rangle \in \Omega  \\
% 0&=h (U_{e}|Y_{s}: a(s) \rightarrow e,U_{f}: f \rightarrow e) \\
% 0&=h (Y_{s}: u \in b(s)| Y_{s'}: s' \rightarrow u, U_{e}: e \rightarrow u), \nonumber\\ %&\text{ \ \ \ \ \ \ \ \ \ \ \ \  \ \ \ \ \ \ \ \ \ \ \ \ \ \ } \forall u \in b(s)\\
% C_{e} &\geq h (U_{e})
\end{align}
for every $\mathcal W \subseteq \mathcal S, \langle \alpha \rangle \in \Omega, e\in \mathcal E, u \in b(s)$ and $s\in\mathcal S$.
Then $\mathcal R \subseteq \mathcal R''$. Replacing $\overline{\Gamma^*}$ by $\Gamma$, %in Definition \ref{def:Improved Outer Bound-vector} 
we obtain an \emph{explicitly computable outer bound} $\mathcal R''(\Gamma)$.

\end{corollary}

%%%%%%%%%%%%%%%%%%%%%%%%%%%%%%%%
\section{Conclusion}
%%%%%%%%%%%%%%%%%%%%%%%%%%%%%%%%

In this paper, we showed that by using auxiliary random variables, entropies are sufficient to uniquely characterise the  probability distribution of a random vector (up to outcome relabelling).
Yet, there are still  many open questions remained to be answered. For example, the number of auxiliary random variables used are exponential to the size of the support. Can we reduce the number of auxiliary random variables? What is the tradeoff between the number of auxiliary variables used and the quality of how well entropies can characterise the distribution? To the extreme, if only one auxiliary random variable can be used, how can one pick the variable to best describe the distribution?

%For example, the proposed construction of the auxiliary random variables are not optimised in any sense.   Suppose we can use only a fixed number of auxiliary random variables, how well  entropies  can represent the correlation among random variables?  This question is still unanswered.

\bibliographystyle{ieeetr}
\bibliography{network}
\begin{appendices}

\section{Scalar case}\label{app.a}

The main ingredients in the proofs for Theorems \ref{thm1} and \ref{thm4} are the properties of the partition random variables, which will be reviewed as follows. 
By understanding the properties, we can better understand the  logic behind Theorem \ref{thm1}. 
%At the same time, many of these properties can also be extended for random vector. These properties are also the main ingredients in the proof for Theorem \ref{thm4}. 

\begin{lemma}[Properties]\label{lemma:secaux1}
Let $X $ be a random variable with support  ${\cal N}_{n}$,  and  $(A_{\lr \alpha \rr}, \: \alpha \in \Omega)$ be its induced binary partition random variables. Then the following properties hold:
\begin{enumerate}
\item (Distinctness) For any $\lr\alpha\rr \neq \lr\beta\rr$, 
\begin{align}
H(A_{\lr \alpha \rr}|A_{\lr \beta \rr})&>0, \label{eq:lemma1a}\\
H(A_{\lr \beta \rr}|A_{\lr \alpha \rr})&>0.\label{eq:lemma1b}
\end{align}

\item (Completetness)
Let $A^{*}$ be a binary random variable such that 
$H(A^{*} | X) = 0$ and $H(A^{*} ) > 0$. Then there exists $\lr\alpha \rr \in \Omega$ such that
\begin{align}
H(A^{*} | A_{\lr \alpha \rr}) = H( A_{\lr \alpha \rr} | A^{*} ) = 0.
\end{align}
In other words, $A_{\lr \alpha \rr}$ and $A^{*}$ are essentially the same.% random variable.

\item (Basis) Let ${\lr \alpha \rr} \in \Omega$. Then there exists 
\[
\lr \beta_{1} \rr , \ldots , \lr \beta_{n-2}  \rr \in \Omega 
\]
such that 
\begin{align}
H(A_{\lr\beta_k\rr}|A_{\lr \alpha \rr},A_{\lr\beta_1\rr},\ldots,A_{\lr\beta_{k-1} \rr})&>0 \label{eq:secAux5}
\end{align}
for all $k=1, \ldots, n-2$. 
%In addition, 
%\begin{align}
%H(A | A_{\lr \alpha \rr},  A_{\beta_1},\ldots,A_{\beta_{n-2-1}}) = 0.  
%\end{align}
\end{enumerate}
\end{lemma}
%\begin{IEEEproof}
%See Appendix \ref{app.A}.
%\end{IEEEproof}

%%%

Among all binary partition random variables, we are particularly interested in those indicator random variables. The following proposition can be interpreted as ``entropic characterisation'' for those indicator random variables.

\begin{proposition}[Characterising indicators]\label{prop:2}
Let $X $ be a random variable of support ${\cal N}_{n}$ where $ n \ge 3$. Consider the binary partition random variables induced by $X$. Then for  all $i \ge 2$,
\begin{enumerate}
\item   
$
H(A_{\lr i\rr}|A_{\lr j \rr}, j > i) >0 
$, and 

\item For all 
$\alpha \in \Omega$ such that   $H(A_{\lr \alpha \rr}|A_{\lr j \rr} , j>i )> 0$,  
\begin{align}
%H(A_{i-1}|A_i,\ldots,A_n) &\leq H(A_{\lr \alpha \rr}|A_i,\ldots,A_n)\label{eq:secAux8}\\
H(A_{\lr i \rr}) &\leq H(A_{\lr \alpha \rr}). \label{eq:secAux9}
\end{align}

\item
Equalities \eqref{eq:secAux9} hold if and only if 
$A_{\lr \alpha \rr}$ is an indicator random variable detecting an element $\ell \in {\cal N}_{n}$  
such that 
$$p_{\ell} = p_{i}.$$  
%
%i) 
%$\alpha \setminus \{i, \ldots, n\} = \{ \ell \}$ for some $ 2 \le \ell \le i-1
%$ such that $p_{\ell} = p_{\ell+1} = \cdots = p_{i-1}$, or ii) 
%$\alpha = [2,n]$ and $p_{1} = p_{2} = \ldots = p_{i-1} $.
%
%$\alpha\setminus [i,n] = [2, i-1]$ and $p_{1} = p_{2} = \cdots = p_{i-1}$.

\item 
Let $\beta \subseteq \{2.\ldots, n\}$. The indicator random variable $A_{\lr 1 \rr}$ is the only binary partition variable of $X$ such that 
\[
H(A_{\lr \alpha \rr} | A_{\lr j \rr}, j \in \beta ) > 0
\]
for all proper subset $\beta$ of $\{2.\ldots, n\}$.
%
%
%Suppose $A_{\lr \alpha \rr}$ is a binary partition random variables such that 
%if and only if $\beta \neq [2,n]$. Then $A_{\lr\alpha\rr}  =A_{\lr 1 \rr}$.
%
\end{enumerate}
\end{proposition}
%\begin{IEEEproof}
%See Appendix \ref{app.A}.
%\end{IEEEproof}

%\begin{remark}
%Note that, in that case, $A_{[2,n]}$ is an indicator variable for $X=1
%$. For this special case, we will also refer $A_{[2,n]}$ to $A_{\lr 1 \rr}$.
%\end{remark}

%%%%%%%%%%%%%%%%%%%%%%%%%%%%%%%%%%%
%\subsection{Distributions characterisation via entropies}
\def\N{{\cal N}}  
%%%%%%%%%%%%%%%%%%%%%%%%%%%%%%%

\begin{IEEEproof}[Sketch of Proof for Theorem \ref{thm1}]
Let $X$  be a random scalar and 
$A_{\lr \alpha \rr}$ for $\lr \alpha \rr\in\Omega$ are its induced partition random variables.  
Suppose  $X^{*}$ is another  random variable  such that 1) the size of its support $\X^{*}$ is at most the same as that of $X$, and 2) there exists random variables $(B_{\lr \alpha \rr}, \alpha \in\Omega)$ satisfying \eqref{eq34} and \eqref{eq35}.

Roughly speaking, \eqref{eq34} and \eqref{eq35} mean that the set of random variables $(B_{\lr \alpha \rr}, \alpha \in\Omega)$ satisfy most properties as ordinary partition random variables. 
To prove the theorem, our first immediate goal is to prove that those random variables $B_{\lr \alpha \rr}$ are indeed binary partition random variables.  In particular, we can prove that 
\begin{enumerate}
\item (\emph{Distinctness}) All the random variables $B_{\lr \alpha \rr}$ for $\lr \alpha \rr\in\Omega$ are distinct and have non-zero entropies.

\item (Basis) Let $\lr\alpha\rr  \in \Omega$. Then there exists 
\[
\lr\beta_{1}\rr , \ldots , \lr\beta_{n-2}\rr \in \Omega 
\]
such that 
\begin{align}
H(B_{\lr\beta_k\rr}|B_{\lr \alpha \rr},B_{\lr\beta_1\rr},\ldots,B_{\lr\beta_{k-1}\rr})&>0  
\end{align}
for all $k=1, \ldots, n-2$. 

\item  (\emph{Binary properties})
For any $\lr\alpha\rr\in\Omega$, $B_{\lr \alpha \rr}$ is a binary partition random variable of $X^{*}$.   In this case,  we may assume without loss of generality that there exists   $\omega_{\lr \alpha \rr} \subseteq \X^{*}$ such that  
\begin{equation} 
  B_{\lr \alpha \rr} = \left\{
  \begin{array}{l l}
    \omega_{\lr \alpha \rr} & \quad \text{if $X^{*} \in \omega_{\lr \alpha \rr}$}\\
    \omega_{\lr \alpha \rr}^{c} & \quad \text{otherwise.}\\
  \end{array} \right.
\end{equation}

\item (Completetness)
Let $B^{*}$ be a binary partition random variable of $X^{*}$ with non-zero entropy. Then there exists $\lr\alpha\rr \in \Omega$ such that
\begin{align}
H(B^{*} | B_{\lr \alpha \rr}) = H(B_{\lr \alpha \rr} | B^{*} ) = 0.
\end{align}
\end{enumerate}

Then by \eqref{eq34} -- \eqref{eq35} and 
Proposition \ref{prop:2}, we show that 
$B_{\lr \alpha \rr}$ satisfies all properties which are only satisfied by the indicator random variables. Thus, we prove that  
$B_{\lr \alpha \rr}$ is an \emph{indicator variable} if  
$|\alpha|=1$. Finally, once we have determined which are the indicator variables, we can immediately determine the probability distribution. 
As $H(A_{\lr \alpha \rr})=H(B_{\lr \alpha \rr})$ for all $\lr \alpha \rr  \in \Omega$, the distribution of $X^{*}$ is indeed  the same as that of $X$ (subject to relabelling).
\end{IEEEproof}
%\begin{IEEEproof}
%See Appendix \ref{app.B}.
%\end{IEEEproof}

%\vspace{-.2cm}

%%%%%%%%%%%%%%%%%%%%%%%%%%%%%%%%
\section{Vector case}\label{app.C}

\def\X{{\cal X}}

\def\bx{{\bf x}}
\def\by{{\bf y}}
\newcommand{\supp}[1]{{\mathcal S}(#1)}
\newcommand\msim[1]{{\: \sim_{#1} \:}}

In this appendix, we will sketch the proof for  Theorem \ref{thm4}, which extends Theorem \ref{thm1} to the random vector case. 

%\subsection{Intermediate results}
Consider a random vector  
\begin{align}
X = (X_m : m \in \N_{M}).
\end{align}
We will only consider the general case\footnote{
In the special case when the support size of $X$ is less than 3, the theorem can be proved directly.
} where the support size of $X$ is at least 3, i.e.,  
$
{\cal S}(X_m : m \in \N_{M}) \ge 3
$.

Let $\X$ be the support of $X$. 
Hence, elements of $\X$ is of the form 
$x=(x_{1},\ldots, x_{M})$ such that 
$$
\Pr(X_{m} = x_{m} , m \in \N_{M}) > 0
$$
if and only if $x \in \X$.

%\begin{remark}
%The notation we used is consistent with the one used in Section \ref{}. In that case, $\alpha$ is chosen to be the subset that do not contain the element $1$ in $\X$.  
%\end{remark}

%
The collection of binary partition random variables induced by the random vector
$X = (X_{m}, m\in \N_{M})$ is again indexed by 
$
(A_{\lr \alpha \rr}, \lr \alpha \rr \in \Omega).
$
As before, we may assume without loss of generality  that  
\begin{equation} 
A_{\lr \alpha \rr} = \left\{
  \begin{array}{l l}
    \alpha & \quad \text{if $X \in \alpha$}\\
    \alpha^{c} & \quad \text{otherwise.}\\
  \end{array} \right.
\end{equation}

Now, suppose  
$
(B_{\lr \alpha \rr} , \lr \alpha \rr \in \Omega)
$ 
is a set of random variables satisfying the properties as specified in Theorem \ref{thm4}.
Invoking Theorem \ref{thm1} (by treating the random vector $X^{*}$ as one discrete variable), we can prove the following.
\begin{enumerate}
\item
The size of the support of $X^{*}$ and  $X$ are the same.

\item
$
B_{\lr \alpha \rr} 
$ 
is a binary partition variable for all $\lr\alpha\rr \in \Omega$.

\item 
The set of variables $(B_{\lr \alpha \rr} , \lr\alpha\rr \in \Omega)$ contains all distinct binary partition random variables induced by $X^{*}$.

\item $B_{\lr x \rr}$ is an indicator variable for all $x\in \X$. 

\end{enumerate}

According to definition,  $A_{\lr x \rr}$  is defined as an indicator variable for detecting $x$. However, while $B_{\lr x \rr}$ is an indicator variable, the subscript $x$ in $B_{\lr x \rr}$ is only an index. The element detected by $B_{\lr x \rr}$ can be any element in the support  of  $X^{*}$, which can be completely different from  $\X$.
To highlight the difference,  we define the mapping $\sigma$  such that 
for any $x\in \X$, $\sigma(x)$ is the element in the support of $X^{*}$ that is detected by $B_{\lr x \rr}$. In other words
\begin{align}
A^{*}_{\lr \sigma(x) \rr} = B_{\lr x \rr}.
\end{align}

The following lemma follows from  Theorem \ref{thm1}. 
\begin{lemma}
For all $x\in \X$,  
$$
\Pr(X=x) = \Pr(X^{*} = \sigma(x)).
$$
\end{lemma}

Let $\X^{*}$ be the support of $X^{*}$. We similarly define 
$\Omega^{*}$ as the collection of all sets of the form $\{ \gamma , \gamma^{c} \}$ where 
$\gamma$ is a subset of $\X^{*}$ and the sizes of $\gamma$ and $\gamma^{c}$ are non-zero. Again, we will use $\lr \gamma \rr$ to denote the set $\{\gamma, \gamma^{c} \}$ and  define  
\begin{equation} 
A^{*}_{\lr \gamma \rr} = \left\{
  \begin{array}{l l}
    \gamma & \quad \text{if $X^{*} \in \gamma$}\\
    \gamma^{c} & \quad \text{otherwise.}\\
  \end{array} \right.
\end{equation}

For any $\lr\alpha\rr  \in \Omega$,  $B_{\lr \alpha \rr}$ is a binary partition random variable of $X^{*}$. Hence, we may assume without loss of generality that there exists $\gamma$ such that 
$
A^{*}_{\lr \gamma \rr} = B_{\lr \alpha \rr}.
$
For notation simplicity, we may further extend\footnote{
Strictly speaking, $\sigma(\alpha)$ is not precisely defined. As $\lr \gamma \rr = \lr \gamma^{c} \rr$, $\sigma(\alpha)$ can either be  $\gamma$ or $\gamma^{c}$. Yet, the precise choice of $\sigma(\alpha)$ does not have any effects on the proof. However, we only require that when $\alpha$ is a singleton, $\sigma(\alpha)$ should also be a singleton.  
} the mapping $\sigma$ such that 
$
A^{*}_{\lr \sigma(\alpha) \rr} = B_{\lr \alpha \rr}
$
for all $ \alpha  \subseteq \X$.

%%%%%%%%%%%%%%%%%%
\begin{proposition}\label{prop2}
Let $\lr \alpha \rr \in \Omega$. Suppose  
$A_{\lr\beta\rr}$ satisfies the following properties:  
\begin{enumerate}
\item For any  $\gamma \subseteq \alpha$, 
$H(A_{\lr \beta \rr} |  A_{\lr x \rr} , x \in \gamma ) = 0$
if and only if $\gamma = \alpha$.

\item For any  $\gamma \subseteq \alpha^{c}$, 
$H(A_{\lr \beta \rr} |  A_{\lr x \rr} , x \in \gamma ) = 0$
if and only if $\gamma = \alpha^{c}$.

\end{enumerate}
Then $A_{\lr \beta \rr} = A_{\lr \alpha \rr}$.
\end{proposition}
\begin{IEEEproof}
Direct verification.
\end{IEEEproof}

By definition of $B_{\lr \alpha \rr}$ and Proposition \ref{prop2}, we have the following 
result.
%%%%%%
\begin{proposition}\label{prop3}
Let $\lr \alpha \rr \in \Omega$. Then  
$B_{\lr \beta \rr} = B_{\lr \alpha \rr}$ is the only binary partition variable of $X^{*}$ such that  
\begin{enumerate}
\item For any  $\gamma \subseteq \alpha$, 
$H(B_{\lr \beta \rr} |  B_{\lr x \rr} , x \in \gamma ) = 0$
if and only if $\gamma = \alpha$.

\item For any  $\gamma \subseteq \alpha^{c}$, 
$H(B_{\lr \beta \rr} |  B_{\lr x \rr} , x \in \gamma ) = 0$
if and only if $\gamma = \alpha^{c}$.

\end{enumerate}
\end{proposition}

%%%%%%%%%%%%%%%%%%
\begin{proposition}\label{prop4}
Let  $\alpha \in \X$. Then 
$\lr \sigma(\alpha) \rr = \lr \delta(\alpha) \rr$, where 
$
\delta(\alpha) = \{ \sigma(x) : x\in\alpha  \}
$. 
%
%
%Let  $\alpha \in \X$ and 
%$
%\delta(\alpha) = \{ \sigma(x) : x\in\alpha  \}
%$. 
%Then  
%\[
%\lr \sigma(\alpha) \rr = \lr \delta(\alpha) \rr.
%\]
\end{proposition}
\begin{IEEEproof}
By Proposition \ref{prop3},   $B_{\lr \alpha \rr} = A^{*}_{\lr \sigma(\alpha) \rr}$    
  is the only variable such that 
\begin{enumerate}
\item For any  $\gamma \subseteq \alpha$, 
$H(A^{*}_{\lr \sigma(\alpha) \rr} |  A^{*}_{\lr \sigma(x) \rr} , x \in \gamma ) = 0$
if and only if $\gamma = \alpha$.

\item For any  $\gamma \subseteq \alpha^{c}$, 
$H(A^{*}_{\lr \sigma(\alpha) \rr} |  A^{*}_{\lr \sigma(x) \rr} , x \in \gamma ) = 0$
if and only if $\gamma = \alpha^{c}$.
\end{enumerate}

The above two properties can then be rephrased as 
\begin{enumerate}
\item For any  $\delta(\gamma) \subseteq \delta(\alpha)$, 
$$
H(A^{*}_{\lr \sigma(\alpha) \rr} |  A^{*}_{\lr \sigma(x) \rr} ,  \sigma(x) \in \delta(\gamma) ) = 0
$$
if and only if $\delta(\gamma) = \delta(\alpha)$

\item For any  $\delta(\gamma) \subseteq  \delta(\alpha^{c})$, 
$$
H(A^{*}_{\lr \sigma(\alpha) \rr} |  A^{*}_{\lr \sigma(x) \rr} , \sigma(x) \in \delta(\gamma) ) = 0
$$
if and only if $\delta(\gamma) = \delta(\alpha^{c})$.
\end{enumerate}
Now, we can invoke Proposition \ref{prop2} and prove that  
$
A^{*}_{\lr \delta(\alpha) \rr} = A^{*}_{\lr \sigma(\alpha) \rr}
$
or equivalently, 
$
\lr\delta(\alpha) \rr = \lr \sigma(\alpha) \rr
$.
The proposition then follows.
\end{IEEEproof}

\begin{proposition}\label{prop5}
Consider two distinct elements $x=(x_{1},\ldots, x_{M})$ and 
$x'=(x'_{1},\ldots, x'_{M})$ in $\X$. 
Let 
\begin{align}
\sigma(x) & =  y = (y_{1}, \ldots, y_{M})  \label{eq:87}\\
\sigma(x') & = y' = (y'_{1}, \ldots, y'_{M}). \label{eq:88} 
\end{align}
Then 
$
x_{m} \neq x'_{m}
$
if and only if 
$
y_{m} \neq y'_{m}
$.
\end{proposition}
\begin{IEEEproof}%[Proof of Proposition \ref{prop5}]
First, we will prove the only-if statement. 
Suppose $x_{m} \neq x'_{m}$. 
 Consider the following two sets
 \begin{align}
 \Delta = \{x''=(x''_{1},\ldots, x''_{M}) \in \X : \: x''_{m} \neq x_{m} \},\\
 \Delta^{c} = \{x''=(x''_{1},\ldots, x''_{M}) \in \X  : \: x''_{m} = x_{m} \}.
 \end{align}
%So, by our assumption, size of $\Delta$ is at least 2. 
It is obvious that 
$
H(A_{\lr\Delta\rr} | X_{m}) = 0.
$
By \eqref{eq34}-\eqref{eq35}, we have  
$
H(B_{\lr\Delta\rr} | X^{*}_{m}) = 0
$.
Hence, 
$
B_{\lr\Delta\rr} = A^{*}_{\lr\sigma(\Delta) \rr}.
$
Since  
$
H(B_{\lr \Delta \rr} | X^{*}_{m}) = 0
$, 
this implies 
$
H( A^{*}_{\lr\sigma(\Delta) \rr} | X^{*}_{m}) = 0
$. 

Now, notice that $x \in \Delta^{c}$ and $x' \in \Delta$.
By Proposition \ref{prop4},  
$
\sigma(\Delta)  =  \{  \sigma(x) : x \in \Delta \}
$. 
 Therefore, 
$
y'=\sigma(x') \in \sigma(\Delta)
$
 and 
$
y=\sigma(x') \not\in \sigma(\Delta).
$
Together with the fact that 
$
H( A^{*}_{\lr\sigma(\Delta) \rr} | X^{*}_{m}) = 0
$, 
we can then prove that 
$$
y'_{m} \neq y''_{m}.
$$

Next, we prove the if-statement. Suppose $y, y' \in \X^{*}$ such that  
$
y_{m} \neq y'_{m}
$. 
There exist  $x$ and $x'$ such that \eqref{eq:87} and \eqref{eq:88} hold. 
Again, define 
 \begin{align}
 \Lambda \defined \{y''=(y''_{1},\ldots, y''_{M}) \in \X^{*} : \: y''_{m} \neq y_{m} \},\\
 \Lambda^{c} \defined \{y''=(y''_{1},\ldots, y''_{M}) \in \X^{*}  : \: y''_{m} = y_{m} \}.
 \end{align}
Then 
$
H(A^{*}_{\lr \Lambda \rr} | X^{*}_{m}) = 0
$.
Let 
$
\Phi \defined \{ x\in \X :\: \sigma(x) \in \Lambda\}.
$
By definition and Proposition \ref{prop4},
$
B_{\lr \Phi \rr}   = A^{*}_{\lr \sigma (\Phi)  \rr}   = A^{*}_{\lr  \Lambda \rr} .
$
Hence, we have $H(B_{\lr\Phi\rr} | X^{*}_{m}) = 0$ and consequently 
$H(A_{\lr\Phi\rr} | X_{m}) = 0$. 
On the other hand, it can be verified from definition that  $x \in \Phi^{c}$ and $x' \in \Phi$. Together with that  
$H(A_{\lr\Phi\rr} | X_{m}) = 0$, we prove that   $x_{m} \neq x'_{m}$. The proposition then follows.
\end{IEEEproof}

%As a corollary of Proposition \ref{prop5},   $\sigma$ maps $\X$ to $\X^{*}$ 
\begin{IEEEproof}[Proof of Theorem \ref{thm4}]
A direct consequence of Proposition \ref{prop5} is that there exists bijective mappings $\sigma_{1} , \ldots, \sigma_{M}$ such that 
$
\sigma(x) = (\sigma_{1}(x_{1}), \ldots, \sigma_{M}(x_{M})).
$
On the other hand, Theorem \ref{thm1} proved that 
$
\Pr( X = x) = \Pr ( X^{*} = \sigma(x)).
$
Consequently, 
\begin{multline}
\Pr(X_{1} = x_{1}, \ldots,  X_{M} = x_{M}) 
\\
= \Pr(X^{*}_{1} = \sigma_{1}(x_{1}), \ldots,  X^{*}_{M} = \sigma_{M}(x_{M})). 
\end{multline}
Therefore,  the joint distributions of $X= (X_{1} , \ldots, X_{M})$ and 
$X^{*}= (X^{*}_{1} , \ldots, X^{*}_{M})$ are essentially the same (by renaming $x_{m}$ as $\sigma_{m}(x_{m})$).
\end{IEEEproof}

\end{appendices}

\end{document}